\documentclass[sigconf]{acmart}
\usepackage{multicol}
\usepackage{multirow}
\usepackage[ruled, lined, linesnumbered, commentsnumbered, longend]{algorithm2e}

\SetCommentSty{mycommfont}

\AtBeginDocument{%
  }

\copyrightyear{2022}
\acmYear{2022}
\setcopyright{rightsretained}
\acmConference[ESEM '22]{ACM / IEEE International Symposium on Empirical Software Engineering and Measurement (ESEM)}{September 19--23, 2022}{Helsinki, Finland}
\acmBooktitle{ACM / IEEE International Symposium on Empirical Software Engineering and Measurement (ESEM) (ESEM '22), September 19--23, 2022, Helsinki, Finland}\acmDOI{10.1145/3544902.3546248}
\acmISBN{978-1-4503-9427-7/22/09}

\begin{document}
\title{Heterogeneous Graph Neural Networks for Software Effort Estimation}

\author{Hung Phan}
\orcid{0000-0001-7464-1597}
\affiliation{%
  \institution{Iowa State University}
  \streetaddress{P.O. Box 1212}
  \city{Ames}
  \state{Iowa}
  \country{USA}
  \postcode{50010}
}
\email{hungphd@iastate.edu}

\author{Ali Jannesari}
\affiliation{%
  \institution{Iowa State University}
  \streetaddress{P.O. Box 1212}
  \city{Ames}
  \state{Iowa}
  \country{USA}
  \postcode{50010}
}
\email{jannesar@iastate.edu}

\renewcommand{\shortauthors}{Phan et al.}

\begin{abstract}
\textbf{Background.} Software effort can be measured by story point \cite{10.1145/2639490.2639503}. Story point estimation is important in software projects' planning. Current approaches for automatically estimating story points focus on applying pre-trained embedding models and deep learning for text regression to solve this problem. These approaches require expensive embedding models and confront challenges that the sequence of text might not be an efficient representation for software issues which can be the combination of text and code.

\textbf{Aims.} We propose HeteroSP, a tool for estimating story points from textual input of Agile software project issues. We select GPT2SP \cite{9732669} and Deep-SE \cite{DBLP:journals/corr/ChoetkiertikulD16} as the baselines for comparison.

\textbf{Method.} First, from the analysis of the story point dataset \cite{DBLP:journals/corr/ChoetkiertikulD16}, we conclude that software issues are actually a mixture of natural language sentences with quoted code snippets and have problems related to large-size vocabulary. Second, we provide a module to normalize the input text including words and code tokens of the software issues. Third, we design an algorithm to convert an input software issue to a graph with different types of nodes and edges. Fourth, we construct a heterogeneous graph neural networks model with the support of fastText \cite{bojanowski2017enriching} for constructing initial node embedding to learn and predict the story points of new issues.

\textbf{Results.} We did the comparison over three scenarios of estimation, including within project, cross-project within the repository, and cross-project cross repository with our baseline approaches. We achieve the average Mean Absolute Error (MAE) as 2.38, 2.61, and 2.63 for three scenarios. We outperform GPT2SP in 2/3 of the scenarios while outperforming Deep-SE in the most challenging scenario with significantly less amount of running time. We also compare our approaches with different homogeneous graph neural network models and the results show that the heterogeneous graph neural networks model outperforms the homogeneous models in story point estimation. For time performance, we achieve about 570 seconds as the time performance in both three processes: node embedding initialization, model construction, and story point estimation. HeterpSP's artifacts are available at \cite{urlHeteroSPArtifact}.

\textbf{Conclusion.} HeteroSP, a heterogeneous graph neural networks model for story point estimation, achieved good accuracy and running time. 
\end{abstract}

\ccsdesc[500]{Software and its engineering}

\keywords{software effort estimation, heterogeneous graph transformer, graph neural networks}

\maketitle

\section{Introduction}
Agile software development (ASD) is the class of software development approaches that can provide the flexibility in delivering products by multiple iterations of development \cite{urlAgileDevelopment}. In ASD, requirements along with solutions are achieved thanks to the collaboration between self-organizing cross-functional teams. A core component of producing the following ASD correctly is the step of planning to make the effort estimation. Software effort estimation is a software engineering (SE) activity that is performed under a development process called planning. There are three common levels of planning in ASD, including release planning, iteration planning, and current-day planning. All of these levels of planning require good effort estimation \cite{10.1145/2639490.2639503}. There are several techniques for effort estimation, including Planning poker, Expert Opinion, Analogy, and Disaggregation \cite{cohn2005agile}. From a study of \cite{10.1145/2639490.2639503}, Planning poker is considered the best Agile planning technique. This approach starts when the product owners or customers went through an Agile user story and described a feature to a group of estimators. Next, each estimator holds planning poker cards with a specific set of integer values such as $[0,1,2,3,5,8..,100]$ that are represented for the measure of how much effort is needed to complete a story. Estimators continue asking questions to product owners and discussing with each other until all estimators unite the assigned effort value for each story.  In ASD, we call that value the story point of a software story \cite{10.1145/2639490.2639503}. In Planning poker, story points are assigned as Fibonacci numbers \cite{6272519}.

Story Points (SP) require a lot of effort from the estimators. To have the information on SPs, estimators need to make multiple panels for discussion to solve the conflict about the assignment of SP for a specific work item. Besides, estimators from specific research domains can have a bias in the decision of SP.  That fact leads to the risk of incorrect story point estimation \cite{10.1145/2639490.2639503}. These inaccurate SPs cause the project to confront the challenges of harmful sprint planning. For example, assigning a story point that is too small compared to the actual effort for completing a work item will cause a reduction in the productivity of sprint development. Besides, the developer time of a story might be much longer than expected, which brings the risk of project failure. Both overestimating stories and underestimating stories cause problems for ASD. Thus, automatically predicting SP given the work item is an important and helpful research problem in SE. It will help to reduce a large amount of effort in the planning process for ASD.

Automatically predicting story points from software stories (issues) in ASD can be formulated as a text regression problem. Given the input as software issues, the output of software effort estimation is the number of efforts estimated. There are several research works that have appeared since 2016 for automatically inferring story points in software engineering \cite{10.1145/2972958.2972959,DBLP:journals/corr/ChoetkiertikulD16,9732669}. The key idea of these works is to suggest the SP in three steps.  First, embedding models are used to learn the vector representation of software issues. Each issue contains a title and a description. Thus, the first task is to learn the embedding of a document with sentences in titles and sentences in descriptions. Second, a training model will be provided by different regression machine learning and deep learning algorithms to learn the SP from the sequence of input vectors from the first step. Third, the model constructed from training will be used to predict story points for new issues. In prior works related to story point estimation, Porru et al. \cite{10.1145/2972958.2972959} used the traditional bag-of-words approach for vectorization and multiple machine learning approaches such as Support Vector Machine, Naive Bayes, and K Nearest Neighbourhood for model construction. Choetkiertikul  et al. \cite{DBLP:journals/corr/ChoetkiertikulD16}  designed a deep learning based approach to solve the problem, called Deep-SE. They use Long Short Term Memory (LSTM) for vectorization and Recurrent Highway Network (RHN) for model construction. Fu et al. \cite{9732669} proposed a story point estimation tool GPT2SP, which relied on a pre-trained language model GPT-2 and Byte Pair Encoding (BPE) tokenization for issue representation. Next, in this approach, a transformer decoder and Multi-Layer Perceptron (MLP) are used for constructing the training model. In general, these works solved the problem by taking advantage of techniques of machine learning and deep learning approaches that are used in Natural Language Processing (NLP).

Prior research works showed that applying NLP techniques in SE problems can be problematic in various ways and cause unexpected low accuracy \cite{DBLP:journals/corr/abs-2003-07914,10.1145/3416506.3423576}. A possible reason is that SE corpora have different characteristics \cite{10.1145/3416506.3423576}. Similarly, in story point estimation, all of Porru's approach \cite{10.1145/2972958.2972959}, Deep-SE \cite{DBLP:journals/corr/ChoetkiertikulD16}, and GPT2SP \cite{9732669} have several drawbacks. First, they rely on a sequence-based learning approach. In other words, they consider the input of the machine learning model as a sequence of text that has a large number of words. This design choice has a disadvantage in that they need to build or rely on a very expensive language model for getting the vector representation of each issue. Deep-SE required 2-8 hours for pre-training to get the vector representation of issues for each single software project in their dataset consisting of 16 projects. However, the replication study conducted by Tawosi et al. \cite{DBLP:journals/corr/abs-2201-05401} showed that the Deep-SE didn't statistically outperform the classical non-neural networks approach TFIDF-SE which is much more faster in model construction. GPT2SP relied on the expensive language model GPT-2 which was trained by the content of  8 million websites in different domains compared to SE. Secondly, another problem related to the running time is that the software issues are actually a combination of source code messages/ source code snippets along with natural language sentences. There has not been any mechanism to differentiate information from code and information from natural language in Deep-SE and GPT2SP due to the limitation of a sequence of text as representation. Another data structure that can substitute sequence-based learning is using graph neural networks (GNN). Phan et al. \cite{https://doi.org/10.48550/arxiv.2203.03062} applied Text Level GNN to predict the range of story points. However, this work also confronted the challenges of the inefficiency of training time, which is 6 times longer than using the traditional Random Forest regression approach.

We attempt to overcome these challenges in this project. We propose HeteroSP, a tool for effectively estimating story points from issues' titles and descriptions. First, to make sure that our approach takes into consideration the characteristic of the software engineering corpus, we do a study of the characteristics of the evaluated dataset of story points proposed by Choetkiertikul et al. \cite{DBLP:journals/corr/ChoetkiertikulD16}. Next, we design an algorithm for constructing the graph from input as software issues' title and description. Third, we construct a heterogeneous graph neural networks (HeteroGNNs) training model to learn information about different types of nodes extracted from software issues. Forth, we use the training model for predicting story points of issues. We propose the following contributions:
\begin{enumerate}
    \item A technique for preprocessing textual information of software issues that highlights the characteristic of software engineering issues.
    \item An algorithm of graph construction from software issues with efficient running time.
    \item A heterogeneous graph neural networks model for story point estimation.
\end{enumerate}
The rest of the paper is provided as follows. In section 2, we introduce the our baseline approaches, GPT2SP \cite{DBLP:journals/corr/ChoetkiertikulD16} and Deep-SE \cite{DBLP:journals/corr/ChoetkiertikulD16}. In section 3, we show an example of the input and output of a software issue. Section 4 describes our study of the dataset to highlight the characteristics of this software engineering corpus. Section 5 illustrates our approach overview and important components. Section 6 mentions the list of research questions. Section 7 shows the results of our approach for each research question. Other sections are Related work, Threats to validity, and Conclusion. 
\section{Background}
\textbf{Story Point Estimation (SPE).} SPE is the main effort estimation approach used in Agile Methodology \cite{cohn2005agile}. In this methodolody, user stories with title and description are provided as an input. Instead of using time for cost estimation, the metric for representing story point is man-hour, which is defined the amount of work completed by an average IT engineer during one hour working continuously \cite{urlManHourDefinition}. The man-hours assigned to each story points are decided by multiple panels of discussion between software project's participants. To our knowledge, GPT2SP \cite{9732669} is the latest tool for automatic SPE while Deep-SE is the most successful approach in the last 5 years until May 2022.


\textbf{GPT2SP.} This tool has four modules for estimating the SP. First, the Subword Tokenization module will tokenize words in each issue to a smaller unit called subword with the use of Byte Pair Encoding (BPE) and the output of GPT-2 language model \cite{radford2018improving}. Second, the Word and Positional Encoding module will init a vector representation with each subword of the input. Third, the Stack of 12 Decoder-only Transformer use a modified version of GPT-2 to generate the representation for software issues. Forth, the Multi-layer Perceptron will be used for regression learning to estimate the story points as real numbers.

\textbf{Deep-SE.} There are two modules of Deep-SE. First, the Document Representation was done by Long Short Term Memory (LSTM) to construct a model for text-to-vector conversion. Second, a Recurrent Highway Network model was constructed for SP estimation \cite{DBLP:journals/corr/ChoetkiertikulD16}. The first module required 2-8 hours for each software project, which can be considered an expensive process \cite{9732669}.

\section{Motivation Example}
\begin{figure}
\centering
\includegraphics[width=\linewidth]{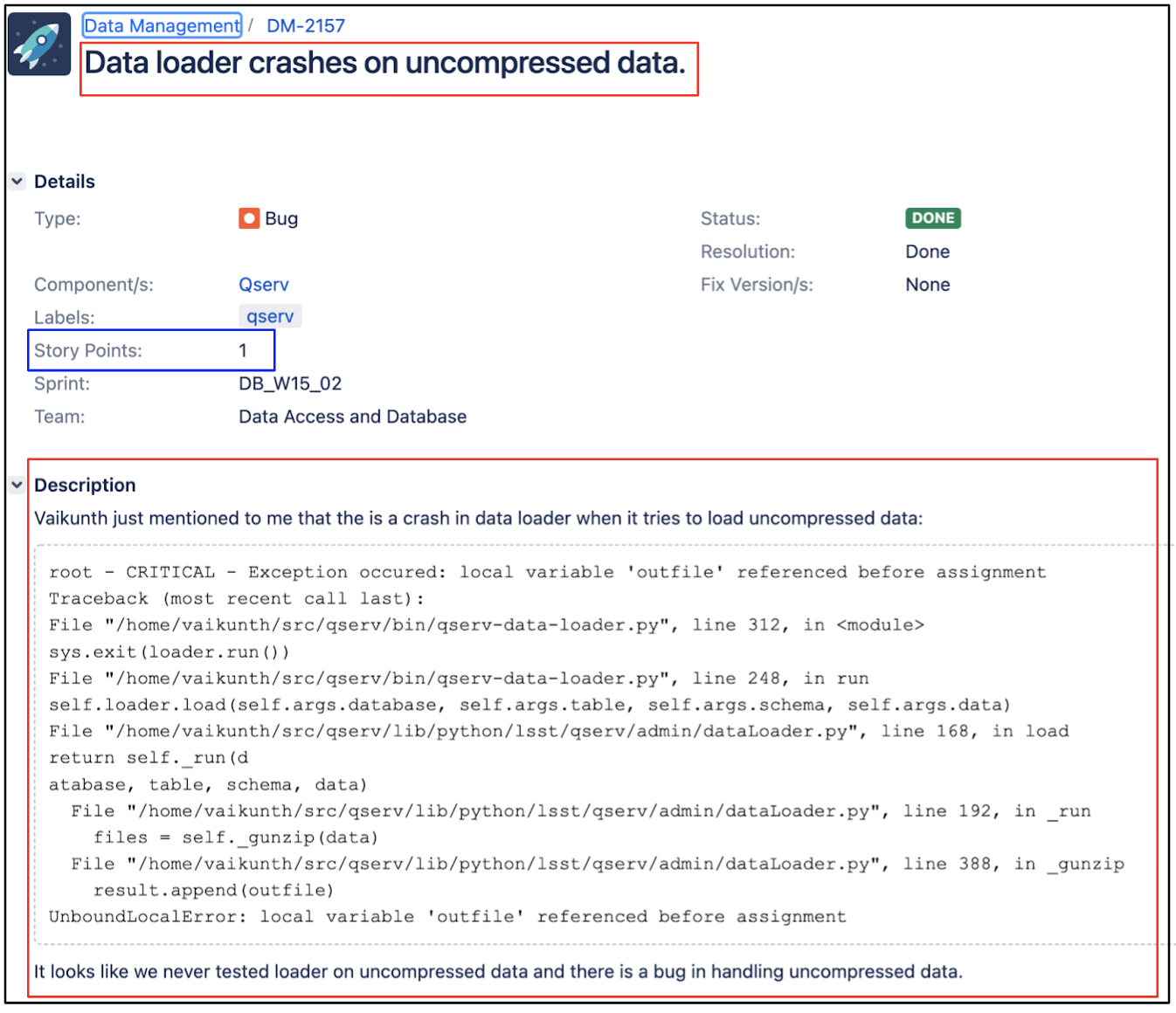}
\caption{Motivation Example: Issue DM-2157 of the Data Management project}
\label{fig:motivationExample}
\end{figure}

An example of a software issue in the Deep-SE dataset is shown in Figure \ref{fig:motivationExample}. From the issue DM-2157\footnote{https://jira.lsstcorp.org/browse/DM-2157} in the Data Management project, we can see that there are two parts of input for story point estimation (highlighted in the red blocks). First, the title part will tell developers about the summarized problem of the issue in one or two sentences. In this example, this sentence mentions the crash of the object called $DataLoader$. Second, the description part will tell the readers about the details of the issue. In this example, the description part contains three components. The first sentence of the description is about a discussion of the issue's creator and one tester. The second part is the information of the compiler returned by the task of the tester. The second sentence is the opinion of the issue's author and some predictions about the reason for the bug. The story point for this issue is 1 (in the blue box), which is the target object for learning and prediction. Title and description are the two inputs provided by the Deep-SE dataset. Other information, such as Fix version, Components, and Sprint id will be omitted from this research problem.

\section{Analysis on Dataset}
Example in Figure \ref{fig:motivationExample} shows that the description of a specific software issue can be a mixture of natural language and non-natural language parts. In this paper, we call the non-natural language part the code part. However, there is a possibility that the example is just an exceptional case in which the issue's creator embeds the code part onto the content of the issue. To test the hypothesis of whether there are a lot of issues containing code parts, we perform an analysis on 23313 issues of the Deep-SE dataset.

Before the analysis, we summarize key points about the Deep-SE dataset. This dataset is collected from 16 software projects available in the Jira system\footnote{https://jira.lsstcorp.org/secure/Dashboard.jspa}. These projects were collected from 8 different repositories. In total there are 23313 issues are collected. The number of issues per project ranges from 384 for the Clover project to 4667 for the Data Management project. The summary of the Deep-SE dataset along with the abbreviation of each project we use in this paper can be seen in Table \ref{tbl:DataStatistics}.

\begin{table}
\small
\centering
\caption{Dataset proposed by Choetkiertikul et al. \cite{DBLP:journals/corr/ChoetkiertikulD16}}
\begin{tabular}{|l|l|l|r|}
\hline
\multicolumn{1}{|c|}{\textbf{Repo.}} & \multicolumn{1}{c|}{\textbf{Project}} & \multicolumn{1}{c|}{\textbf{Abb.}} & \multicolumn{1}{c|}{\textbf{\# issues}} \\ \hline
\multirow{2}{*}{Apache}              & Memos                                 & ME                                 & 1680                                    \\ \cline{2-4} 
                                     & Usergrid                              & UG                                 & 482                                     \\ \hline
\multirow{3}{*}{Appeclerator}        & Appecelerator Studio                  & AS                                 & 2919                                    \\ \cline{2-4} 
                                     & Aptana Studio                         & AP                                 & 829                                     \\ \cline{2-4} 
                                     & Titanium                              & TI                                 & 2251                                    \\ \hline
DuraSpace                            & DuraCloud                             & DC                                 & 666                                     \\ \hline
\multirow{3}{*}{Atlassian}           & Bamboo                                & BB                                 & 521                                     \\ \cline{2-4} 
                                     & Clover                                & CV                                 & 384                                     \\ \cline{2-4} 
                                     & JIRA Software                         & JI                                 & 352                                     \\ \hline
Moodle                               & Moodle                                & MD                                 & 1166                                    \\ \hline
Lsstcorp                             & Data Management                       & DM                                 & 4667                                    \\ \hline
\multirow{2}{*}{Mulesoft}            & Mule                                  & MU                                 & 889                                     \\ \cline{2-4} 
                                     & Mule Studio                           & MS                                 & 732                                     \\ \hline
Spring                               & Spring XD                             & XD                                 & 3526                                    \\ \hline
\multirow{2}{*}{Talendforge}         & Talend Data Quality                   & TD                                 & 1381                                    \\ \cline{2-4} 
                                     & Talend ESB                            & TE                                 & 868                                     \\ \hline
Total                                &                                       &                                    & 23313                                   \\ \hline
\end{tabular}
\label{tbl:DataStatistics}
\end{table}

\textbf{Analysis of the size of the vocabulary.} We count the number of unique words per project. In this configuration, we use the raw text of the title and description for each issue as the input for our analysis. 

\textbf{Analysis of average appearance in issue (Avg. App.).} We consider another metric to evaluate the sparsity of the vocabulary in the Deep-SE dataset along with the vocabulary as the average appearance of a word in issues.  

\textbf{Analysis of the number of code parts.} We count the number of code parts that appeared in the issues. We detect the code part by the regular expression on each word in the combination of title and description.

The results of analyzing the size of vocabulary and Avg. App. are shown in the preprocess column of Table \ref{tbl:Preprocess}.  Most of the software projects have a vocabulary size greater than 5000. Besides, the Avg. App for projects is relatively low. They are from 2.77 for Bamboo project to 6.59 for Appcelerator Studio (AS) project. It means that for a general word that appeared in the Bamboo dataset, there are only three issues that contained that word. 

The summary of popular special tags along with a number of issues is shown in Table \ref{tbl:SpecialTag}. We can see that there are a lot of issues that have special tags in the content. The tag $\{noformat\}$ is used to split the content of compilation messages. About one-third of 23313 issues have special tags in their content. It proves that the combination of code parts and natural language parts is common in the story point dataset Deep-SE.

\textbf{A good story point estimation approach should handle the problem of large size vocabulary and preserve the information of software issues as the combination of natural language parts and code parts in its model.}

\begin{table}
\small
\centering
\caption{Statistic on Special Tags of Deep-SE dataset \cite{DBLP:journals/corr/ChoetkiertikulD16}}
\begin{tabular}{|l|r|}
\hline
\multicolumn{1}{|c|}{\textbf{Special Tag}} & \multicolumn{1}{c|}{\textbf{\# of issues}} \\ \hline
\{code\}                                   & 6371                                       \\ \hline
\{noformat\}                               & 1069                                       \\ \hline
\{quote\}                                  & 333                                        \\ \hline
\{code:java\}                              & 182                                        \\ \hline
\{code:javascript\}                        & 178                                        \\ \hline
\textless{}redacted\textgreater{}          & 118                                        \\ \hline
\{code:xml\}                               & 107                                        \\ \hline
\end{tabular}
\label{tbl:SpecialTag}
\end{table}

\begin{table}
\small
\caption{Impact of pre-processing input description of issues}
\centering
\begin{tabular}{|l|rr|rr|}
\hline
\multicolumn{1}{|c|}{\textbf{Proj.}} & \multicolumn{2}{c|}{\textbf{Before Preprocess}}             & \multicolumn{2}{c|}{\textbf{After Preprocess}}              \\ \hline
                                     & \multicolumn{1}{l|}{Vocab} & \multicolumn{1}{l|}{Avg. App.} & \multicolumn{1}{l|}{Vocab} & \multicolumn{1}{l|}{Avg. App.} \\ \hline
AS                                   & \multicolumn{1}{r|}{26301} & 6.59                           & \multicolumn{1}{r|}{23580} & 7.64                           \\ \hline
AP                                   & \multicolumn{1}{r|}{13424} & 3.79                           & \multicolumn{1}{r|}{12152} & 4.50                           \\ \hline
BB                                   & \multicolumn{1}{r|}{7946}  & 2.77                           & \multicolumn{1}{r|}{7688}  & 3.31                           \\ \hline
CV                                   & \multicolumn{1}{r|}{7906}  & 2.85                           & \multicolumn{1}{r|}{7396}  & 3.26                           \\ \hline
DM                                   & \multicolumn{1}{r|}{32617} & 5.63                           & \multicolumn{1}{r|}{25516} & 7.18                           \\ \hline
DC                                   & \multicolumn{1}{r|}{6699}  & 4.60                           & \multicolumn{1}{r|}{4915}  & 6.02                           \\ \hline
JI                                   & \multicolumn{1}{r|}{4471}  & 4.63                           & \multicolumn{1}{r|}{3337}  & 5.86                           \\ \hline
ME                                   & \multicolumn{1}{r|}{37242} & 3.34                           & \multicolumn{1}{r|}{34123} & 3.64                           \\ \hline
MD                                   & \multicolumn{1}{r|}{14236} & 4.35                           & \multicolumn{1}{r|}{11034} & 5.36                           \\ \hline
MU                                   & \multicolumn{1}{r|}{9769}  & 4.14                           & \multicolumn{1}{r|}{8386}  & 5.00                           \\ \hline
MS                                   & \multicolumn{1}{r|}{7101}  & 4.17                           & \multicolumn{1}{r|}{6430}  & 4.87                           \\ \hline
XD                                   & \multicolumn{1}{r|}{26544} & 5.43                           & \multicolumn{1}{r|}{23105} & 6.78                           \\ \hline
TD                                   & \multicolumn{1}{r|}{13257} & 4.50                           & \multicolumn{1}{r|}{12532} & 5.55                           \\ \hline
TE                                   & \multicolumn{1}{r|}{13394} & 3.77                           & \multicolumn{1}{r|}{11913} & 4.53                           \\ \hline
TI                                   & \multicolumn{1}{r|}{41839} & 4.15                           & \multicolumn{1}{r|}{36038} & 5.20                           \\ \hline
UG                                   & \multicolumn{1}{r|}{5628}  & 3.09                           & \multicolumn{1}{r|}{5254}  & 3.84                           \\ \hline
\end{tabular}
\label{tbl:Preprocess}
\end{table}

\section{Approach}
In this section, we present HeteroSP, a heterogeneous graph neural networks model for story point estimation.
\begin{figure*}
\centering
\includegraphics[width=0.9\linewidth]{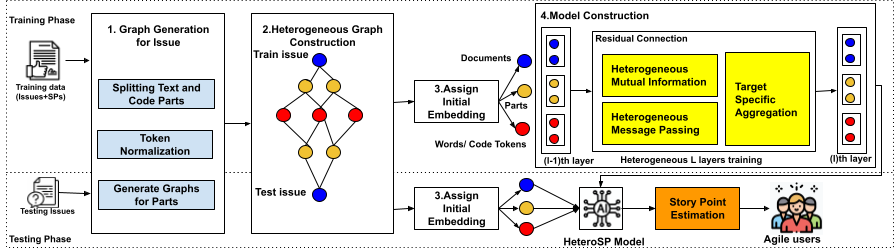}
\caption{An Overview Architecture of HeteroSP}
\label{fig:overviewArchitecture}
\end{figure*}

\textbf{Overview:} Given the software issue as the input, a graph with different types of nodes and edges for each issue is constructed. Second, a heterogeneous graph generated by the combination of multiple graphs for software issues is created. Third, each node in the heterogeneous graph has the initial embedding based on the textual content of their leaf nodes. Fourth, the model building process is done by multiple layers of heterogeneous graph transformer (HGT) training. Fifth, the trained model is used for prediction from the graph extracted from the testing issue to generate the story point for Agile users. The heart of HeteroSP is the engine of heterogeneous graph transformer (HGT) which we apply and optimize from the work of Fey et al. \cite{fey2019fast}. Compared to other traditional GNN models such as TextLevelGNN \cite{https://doi.org/10.48550/arxiv.2203.03062}, HGT has been considered advantageous since this model allows different types of nodes and edges for model construction.  We describe components of the HeteroSP below.
\subsection{Graph Generation for Issue}
\subsubsection{Types of nodes in Graph}
\begin{figure}
\centering
\includegraphics[width=0.8\linewidth]{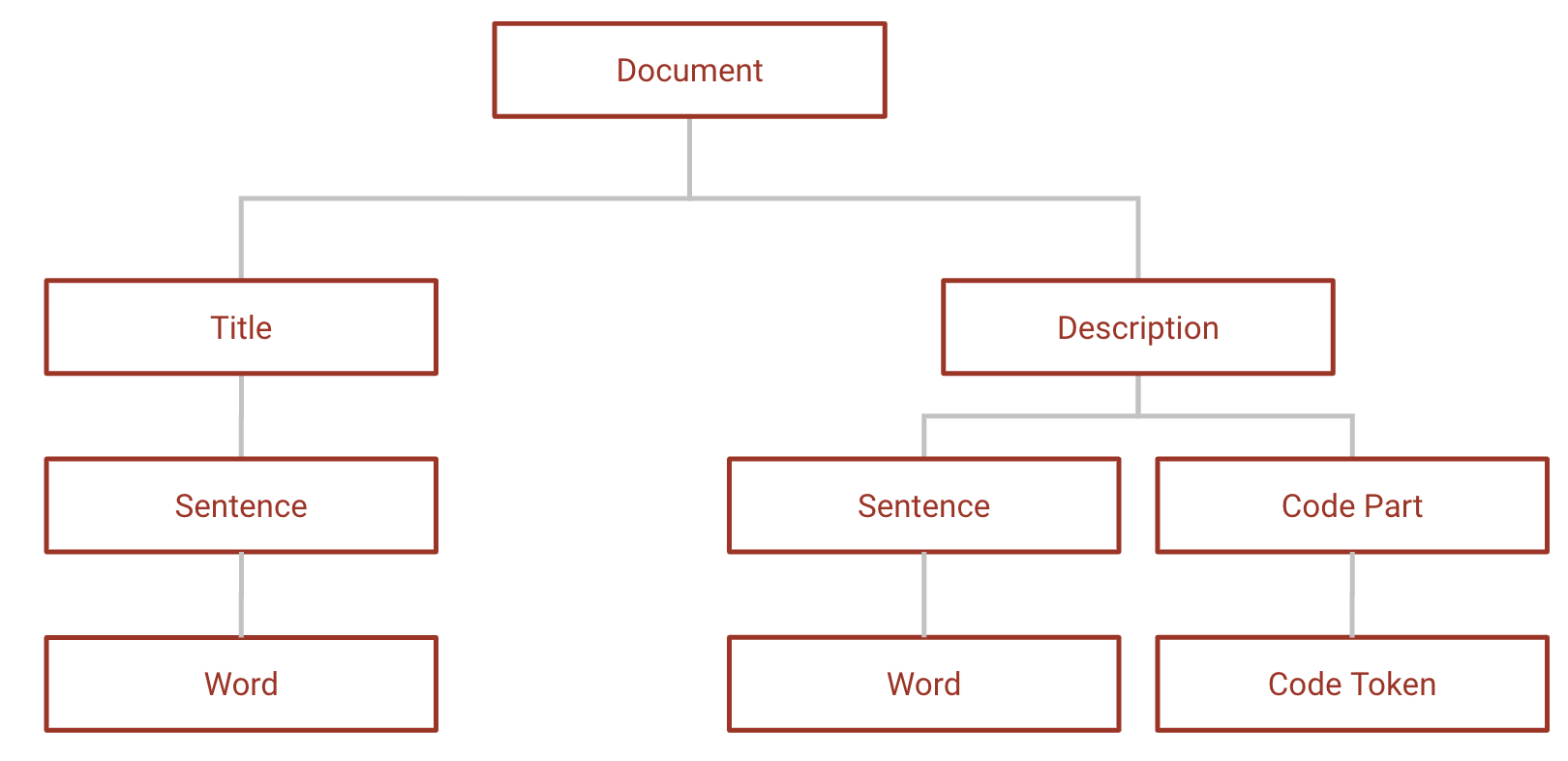}
\caption{Types of nodes in HeteroSP}
\label{fig:fig:typesOfNode}
\end{figure}

From the analysis in the previous section, we design the list of types for nodes for our constructed graph from the textual description as in Figure \ref{fig:fig:typesOfNode}. Each issue will have a root node of type Document. The children of the Document are Title and Description nodes, which are the corresponding summarized nodes for the title and description of the issue. The Title node can be the parent node of multiple Sentence nodes. Each sentence will contain a sequence of words. Compared to the Title node, the Description node can accept a child as Code Part, which is the root node of all code tokens that appeared inside special tags shown in Table \ref{tbl:SpecialTag}.

\subsubsection{Algorithm for Graph Construction}
\begin{algorithm}
\small
    \SetKwFunction{isOddNumber}{isOddNumber}
    \SetKwFunction{splitToParts}{splitToParts}
    \SetKwFunction{createGraphFromParts}{createGraphFromParts}
    \SetKwInOut{KwIn}{Input}
    \SetKwInOut{KwOut}{Output}
    \KwIn{title,description: string}
    \KwOut{nodeDocument: Node}
    
    $nodeTitle\gets new\ Node()$
    
    $nodeDesc\gets new\ Node()$
    
    $listPartsTitle \gets \splitToParts(title)$
    
    $listPartsDesc \gets \splitToParts(description)$
    
    $\createGraphFromParts(nodeTitle,listPartsTitle)$
    
    $\createGraphFromParts(nodeDesc,listPartsDesc)$
    
    $nodeDocument\gets new\ Node()$
    
    $nodeDocument.apppendChild(nodeTitle)$
    
    $nodeDocument.apppendChild(nodeDesc)$

    \KwRet{$nodeDocument$}
    \caption{Graph Construction from Software Issue}
    \label{algm:GraphConstruction}
\end{algorithm}
The algorithm for generating a graph for each software issue is shown in Algorithm \ref{algm:GraphConstruction}. The content of the title and description will be analyzed and split into smaller parts in lines 3 and 4. The construction of graphs for the Title node and Description node is done by lines 5 and 6. Finally, the Document node will be created as the root of the Title and Document node. We explain the details of each function below.

\textbf{Purpose of $splitToParts$ function.} This function will split the content of the title by stop words in English and split the content of the description by stop words along with the special tags defined in Table \ref{tbl:SpecialTag}. The output of this function is multiple parts that can be sentences or code parts.

\textbf{Purpose of $createGraphFromParts$ function.} This function goes over each part as a loop. In each iteration, it creates the graph for each part by traversing each token inside the part and doing the preprocessing for each token. The token after preprocessing is added as Word nodes.

\textbf{Token Normalization}. The preprocessing of the token has the following steps. First, the token is split by camel cases into multiple subtokens. Second, for each subtokens, we perform stemming and lemmatization to have the final tokens. We use StanfordCoreNLP toolkit \cite{manning2014stanford} for text normalization. The text normalization reduces the size of vocabulary, along with increasing the average appearance of a token in the dataset as shown in the After Preprocess column in Table \ref{tbl:Preprocess}. 

\subsubsection{Example of Graph Representation for Issues}
\begin{figure}
\centering
\includegraphics[width=\linewidth]{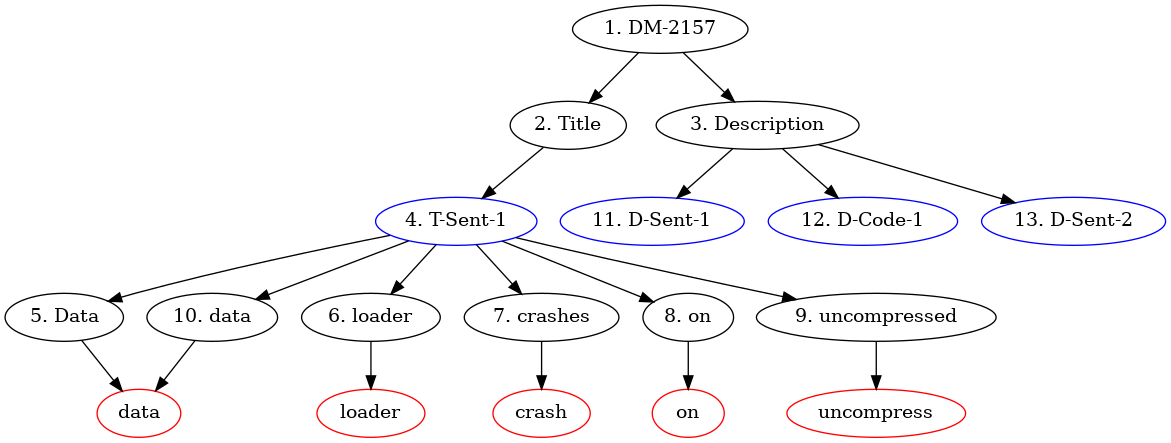}
\caption{Sub-graph extracted of Issue DM-2157 in Figure \ref{fig:motivationExample}}
\label{fig:graphRepresentationExample}
\end{figure}

Part of the graph constructed for issue DM-2157 in Figure \ref{fig:motivationExample} can be shown in Figure \ref{fig:graphRepresentationExample}. In this example, 4 parts of the issues, including one sentence from the title and 2 sentences plus one code part in the implementation will be represented on the graph. The terminal nodes will be Word nodes which are normalized. 

\subsection{Heterogeneous Graph Construction}
Multiple graphs for each issue will be combined into a heterogeneous graph in this step. There are nodes that are unique for each issue and there are nodes that are generalized to shared nodes between issues. The shared nodes (red nodes in Figure \ref{fig:motivationExample}) helped the model to avoid having disjoint sub-graphs in the heterogeneous graph, which can badly impact the training \cite{fey2019fast}. In HeteroSP, we design to split the independent nodes and shared nodes between issues by types of nodes. We select nodes of ancestor types in Figure \ref{fig:fig:typesOfNode} as the independent nodes and nodes of type Word or CodeToken are shared nodes. In a heterogeneous graph, the independent nodes will be assigned their identities by concatenating the issue identification and value of nodes in the individual graphs. The shared nodes will have the identities as the value of nodes in graphs for each issue. An example of graph in Figure \ref{fig:graphRepresentationExample}, non terminal nodes such as \textit{T-Sent-1} will be unique in the heterogeneous graph. The terminal nodes like \textit{"data"} will be generalized and can be connected with nodes of other issues that have  \textit{"data"} in their content.
\subsection{Assign Initial Embedding}
Each node will have an initial vector representation before model construction. Due to the characteristic of the SE corpus, we have to use an initial embedding model that supports embedding for subwords and is also efficient in terms of running time. We apply the fastest \cite{bojanowski2017enriching} as the embedding model for this task. We choose the \textbf{cbow} model with 100 dimensions as the configuration for assigning initial embedding. The embedding of non-terminal nodes like Sentence nodes and Document nodes will be the vector representation of a sequence of text that they covered. The embedding of terminal nodes such as Word nodes will be the vector represented for their nodes' values. 
\subsection{Model Construction}
We inherit the Heterogeneous GNN model proposed by Fey et al \cite{fey2019fast} for training the model for story point estimation. The idea of this engine is to create different layers for learning in different types of nodes. In Figure \ref{fig:overviewArchitecture}, there are three sub-layers related to three types of nodes: Document, Sentence, and Word. A heterogeneous graph transformer will provide the learning process with \textit{L} layers. \textit{In our work, we choose the number of layers L as 2}. Information between each sub-layers will be connected by the residual connection module, which has three components. The first component is heterogeneous mutual information, which is used to assign a weight for more important neighbor nodes in different distributions. The second component, heterogeneous message passing, is used to pass the information between neighbor nodes. The third component, target-specific aggregation, is used to combine the calculation based on the message passing and mutual information to pass the information to the next layer. The implementation of residual connection for each layer of HGT is implemented by Hu et al. \cite{DBLP:journals/corr/abs-2003-01332}.

Since Fey et al. \cite{fey2019fast} designed their model for the classification of the category of authors and papers in a dataset of research papers in NLP, we adjust the library for estimating story points. First, we formulate the types of nodes for prediction as Document nodes. The reason for this design selection is that Document nodes are considered the root nodes for each issue. Second, we substitute the final layer of the HGT engine from Fey et al. \cite{fey2019fast} to a linear transformation layer, which outputs a numeric value instead of a vector with $n$ dimensions ($n$ is the number of classes for classification). Third, while Fey et al. \cite{fey2019fast} used cross-entropy as a loss function, we substitute this loss function to \textit{$L1$} loss function \cite{urlL1Loss} to make our model suitable for calculating the Mean Absolute Error per each step of training. 
\section{Research Questions}
We attempt to answer the following research questions (RQs):
\begin{enumerate}
    \item RQ1: How well can HeteroSP perform in within project effort estimation?
    \item RQ2: How well can HeteroSP perform in cross-project within repository effort estimation?
    \item RQ3: How well can HeteroSP perform in cross-project cross repository effort estimation?
    \item RQ4: Can the proposed graph structure support homogeneous graph neural networks effort estimation?
    \item RQ5: How well can HeteroSP perform in issues without title or without description?
    \item RQ6: How well can HeteroSP perform in story point classification?
    \item RQ7: How well can statistical NLP parse trees help for effort estimation?
\end{enumerate}
We describe the purposes of each RQs along with how we set up the experiments below.
\subsection{Setup}
\textbf{Configuration. } We train our HGT model with 128 hidden channels. We use four attention heads for mutual attention learning. We construct two HGT convolutional layers for each type of node in our heterogeneous graph. For each RQs, we run our model 10 times and take the average MAE as the final results. The summary of our configuration can be shown in Table \ref{tbl:Config}.
\begin{table}
\small
\centering
\caption{Configuration of HeteroSP}
\begin{tabular}{|l|r|}
\hline
\multicolumn{1}{|c|}{\textbf{Config}} & \multicolumn{1}{c|}{\textbf{Num.}} \\ \hline
Attention Head                        & 4                                  \\ \hline
Epoch                                 & 500                                \\ \hline
Convolutional Layer                   & 2                                  \\ \hline
Hidden Channels                       & 128                                \\ \hline
\end{tabular}
\label{tbl:Config}
\end{table}

\textbf{State-of-the-art Approaches.} We compare HeteroSP with the tools GPT2SP \cite{9732669} and Deep-SE \cite{DBLP:journals/corr/ChoetkiertikulD16}. We provide the head-to-head comparison with GPT2SP and Deep-SE from RQ1 to RQ3 per each project. We report the MAE retrieved from the result of the GPT2SP replication package with the last commit at March 10\textsuperscript{th}, 2022. The raw prediction results for each issues of GPT2SP are available at this site\footnote{https://tinyurl.com/mu7jpbur}. We observe that for RQ1, the average MAE provided by their \textbf{testing\_results} folder is 2.57, which is higher than what they reported from their diagram\footnote{https://tinyurl.com/mwhmmdnw} as 1.44. We contacted the author and they suggest to use the results in their diagram for RQ1. We use the results reported in the \textbf{testing\_results} folder for RQ2 and RQ3. Similarly, we use the results from 3 RQs reported from Deep-SE \cite{DBLP:journals/corr/ChoetkiertikulD16} for comparison.

\subsection{RQ1: How well can HeteroSP perform in within project effort estimation?}
In this RQ, we want to test the ability of HeteroSP to predict the story point based on previous planning and estimation of software issues in the past. In this scenario, for each project, we use a set of software issues that appeared earlier in the dataset as training data. The validation and testing issues are the most recent issues that appeared later compared to issues in the training dataset. As this is within project configuration, we set up 16 models for training and evaluation for 16 software projects. We use 60\% of data for training, 20\% of data for validation, and 20\% of data for testing. Our configuration is the same as the experiment provided in GPT2SP.

\textbf{Metric for evaluation.} Similar to Deep-SE and GPT2SP, we use Mean Absolute Error (MAE) to calculate the accuracy of effort estimation. The MAE calculated the average distance between the expected story point and the predicted story point for each issue. MAE is a traditional metric used for regression problems. It has been used as the evaluation metric for all research works related to story point estimation in Software Engineering \cite{DBLP:journals/corr/ChoetkiertikulD16,9732669}.
\subsection{RQ2: How well can HeteroSP perform in cross-project within repository effort estimation?}
In this RQ, we want to set up the scenario that a group of software developers in a software organization want to develop a new software project. Can they rely on the software issues that were made in the development of previous projects developed by their counterparts in the same organization? From this scenario, developers will save time for requirements and planning thanks to the material of the organization in the past.

We follow the same configuration with GPT2SP to set up this experiment. We provide 8 training and testing processes. For each process, we use all the story points from one project for training. The constructed model will be used to predict all the story points of the target projects. There are 8 pairs of source and target projects.
\subsection{RQ3: How well can HeteroSP perform in cross-project cross repository effort estimation?}
The purpose of this research question is to simulate a practical challenge in Jira development \cite{DBLP:journals/corr/ChoetkiertikulD16}. In some projects, there are not enough issues for training the model for estimation. Even other projects from the same organization might have a lack issues to produce an efficient training model. In this case, we need to use the story issues from other projects that are in different repositories to enrich the data for story point estimation.

We select the same source and target projects for evaluation with GPT2SP \cite{9732669} in our experiment. They are AS-MU, AS-MS, CV-UG, MS-TI, MU-TI, AD-AS, TD-AP and TE-ME. Since this configuration is considered the most challenging task in three RQs from the work of Deep-SE \cite{DBLP:journals/corr/ChoetkiertikulD16} and GPT2SP \cite{9732669}, we also use this configuration for experiments of our next RQs, which are more related to intrinsic evaluation about the efficiency of our constructed model.
\subsection{RQ4: Can HeteroSP's graphs support homogeneous GNN in effort estimation?}
Phan et al. \cite{https://doi.org/10.48550/arxiv.2203.03062} is the first work that has attempted to apply Graph Neural Networks classification for predicting the range of story points of the Deep-SE dataset. They rely on the graph construction and training process of the TextLevelGNN engine \cite{DBLP:journals/corr/abs-1910-02356} to solve the problem of estimation. Although TextLevelGNN is a well-known homogeneous GNN model in NLP research, Phan et al. \cite{https://doi.org/10.48550/arxiv.2203.03062} show that the original GNN models can cause negative impacts on estimating story points due to the large size of vocabulary and number of edges for graph construction.

In this RQ, we want to test whether our algorithm \ref{algm:GraphConstruction} for graph construction can achieve better performance for not only heterogeneous GNN but also homogeneous GNN models. The main difference between heterogeneous GNN models and homogeneous GNN models is that the homogeneous GNN models accept only a unique type of nodes and edges as the input. 

We did not use TextLevelGNN \cite{DBLP:journals/corr/abs-1910-02356} approach since they use the n-grams strategy in NLP for graph construction which is inefficient in story point estimation. Instead, we select three well-known homogeneous GNN models that allow customizing the graph construction process. They are Graph Convolutional Networks (GCN) \cite{DBLP:journals/corr/KipfW16}, Graph Attention Networks (GAT) \cite{GATvelickovic2018graph} and Attention-based Graph Neural Networks (AGNN) \cite{AGNNthekumparampil2018attentionbased}. We use the same configuration defined in Table \ref{tbl:Config} to set up the experiment for this RQ. We run these three GNN models with the cross projects cross repositories  estimation.
\subsection{RQ5: How well can HeteroSP perform in issues without title or without description?}
In this RQ, we want to test the ability of HeteroSP on incomplete issues. We simulate the practical case where the issue's creator has already named the issue but hasn't written the description yet. Besides, there is another case where developers forgot to set the title of the issue. To set up the experiment, we performed the training processes of cross-project cross repository effort estimation from two types of input: issues with only title and issues with only description.
\subsection{RQ6: How well can HeteroSP perform in story point classification?}
The output of the story point can range from 1 to 100 in the Deep-SE dataset. However, story points are usually assigned as Fibonacci numbers \cite{DBLP:journals/corr/ChoetkiertikulD16}. The specific set of Fibonacci numbers from 0 to 100 is smaller than the set of integers from 0 to 100. This fact prompted us to think about the hypothesis that a regression model can be replaced by a classification model. In this model, the value of the story point will be assigned as the label for multi-label classification.

To conduct the experiment for this RQ, we changed the HGT model to a node classification model. We changed the \textit{L1} loss function in regression training to a cross-entropy loss function. We performed the experiment on cross-project cross repository estimation.

\subsection{RQ7: How well can statistical NLP parse trees help for Story Point Estimation?}
\begin{figure}
\centering
\includegraphics[width=0.45\linewidth]{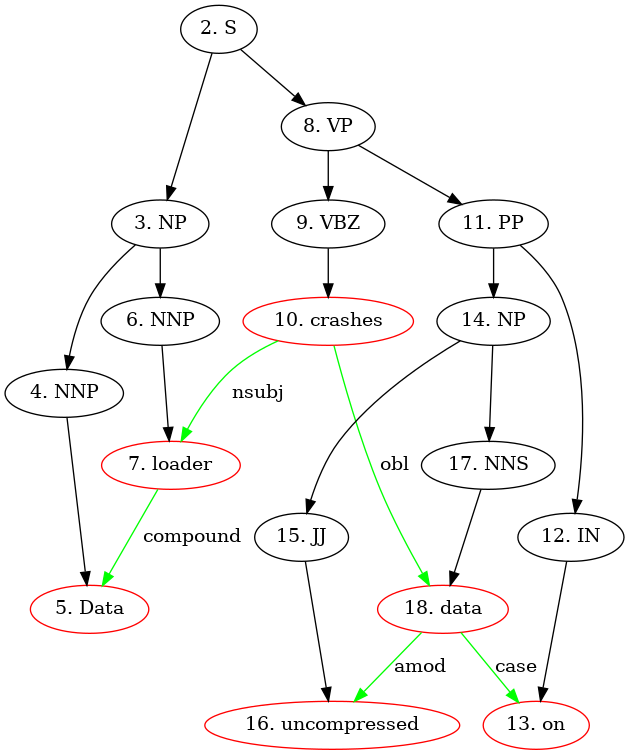}
\caption{The sub-graph of parse tree of issue DM-2157}
\label{fig:examplePOS}
\end{figure}

There has been some recent research on generating graphs from natural language. In these works, StanfordCoreNLP \cite{manning2014stanford} provided an approach to generate a parse tree (or Part-of-Speech POS tree) along with universal dependencies edges. Since the parse tree is well-known in NLP research, we wonder if it can make the impact of using the parse tree to replace our graph representation. An example of a parse tree with universal dependencies edges is shown in Figure \ref{fig:examplePOS}. From our knowledge, there are advantages and disadvantages of replacing our graph construction mechanism with parse tree construction.

\textbf{Advantages.} The parse tree generated from StanfordCoreNLP provides useful information related to the parse tags for each sentence. Besides, there is extra information between nodes that were represented by universal dependencies edges.

\textbf{Disadvantages.} We observe two disadvantages of using a parse tree for constructing a model for HeteroSP. First, the cost of parse tree generation is high. There are eight hours of running the StanfordCoreNLP model on a server with 64 GB of RAM memory and a core-i9 processor to generate a parse tree for 23313 software issues of the Deep-SE dataset. Second, the StanfordCoreNLP parse tree might not be suitable for SE corpora.

We tested our hypothesis of the pros and cons of using the parse tree by this experiment. We changed the graph constructed by our algorithm to the information of the parse tree to construct a new HGT model. Since embedding the whole parse tree can cause exponential increase of nodes and edges, we embedded information of terminal nodes and their directly ancestor nodes for graph construction. Next, we ran the effort estimation on cross-project cross repositories configuration.
\section{Results}
\subsection{RQ1: Within project effort estimation}
We summarize the result of within project estimation in Table \ref{tbl:RQ1}. We achieved better MAE than Deep-SE on the AS project and Moodle project, while GPT2SP achieve the best MAE over 16 projects. A possible reason is due to the fact that the GNN models are usually hungry for data \cite{urlGNNDataHungry} while within repository estimation learned the model from the small number of the training set. We achieved the best MAE for project Bamboo with 0.75 MAE. A possible reason for the lower MAE of Bamboo compared to the other projects is that the estimated story points are consistent over the history of project planning in this project. Besides, the range of the minimum story point to maximum story point is from 1 to 20, which means that the space of numbers assigned for issues of the Bamboo project is more continuous than other projects such as Moodle.

We get the highest MAE with the Data Management project. This project caused challenges to machine learning models in predicting story points. A possible reason is that the training model predicts incorrectly for issues with high story points assigned. The Moodle project is another challenging project with MAE achieved by HeteroSP as 5.34. We outperformed Deep-SE for this project. HeteroSP and GPT2SP achieved about the same accuracy as the Bamboo project. On average, we achieved 2.38 MAE over 16 projects, while GPT2SP achieved 1.44 and Deep-SE achieved 2.09.

We also analyze the performance of HeteroSP in terms of running time. We achieve the total running time as \textbf{18.47} seconds over 4671 issues for this configuration. Since other deep learning models such as Deep-SE required 2-8 hours of training per project, our approach shows the benefits of learning with more representative data structures such as graphs. The data management project required most of the time for running with over 21 seconds. All projects have sufficient testing time from 0.01 seconds to 0.23 seconds.

\textbf{HeteroSP achieves the average MAE as 2.38, which has a higher average MAE than GPT2SP and Deep-SE for within project effort estimation.}
\begin{table}
\small
\centering
\caption{RQ1: Comparison of MAE between HeteroSP and baseline approaches for Within Project Estimation}
\begin{tabular}{|ll|r|r|r|}
\hline
\multicolumn{1}{|c|}{\textbf{Project}}    & \multicolumn{1}{c|}{\textbf{Abbrev}} & \multicolumn{1}{c|}{\textbf{HeteroSP}} & \multicolumn{1}{c|}{\textbf{GPT2SP}} & \multicolumn{1}{c|}{\textbf{Deep-SE}} \\ \hline
\multicolumn{1}{|l|}{Appcelerator Studio} & AS                                   & 1.30                                   & 0.84                                 & 1.36                                  \\ \hline
\multicolumn{1}{|l|}{Aptana Studio}       & AP                                   & 3.24                                   & 1.93                                 & 2.71                                  \\ \hline
\multicolumn{1}{|l|}{bamboo}              & BB                                   & 0.75                                   & 0.44                                 & 0.74                                  \\ \hline
\multicolumn{1}{|l|}{Clover}              & CV                                   & 3.64                                   & 1.98                                 & 2.11                                  \\ \hline
\multicolumn{1}{|l|}{Data Management}     & DM                                   & 6.19                                   & 3.10                                 & 3.77                                  \\ \hline
\multicolumn{1}{|l|}{Duracloud.csv}       & DC                                   & 0.78                                   & 0.48                                 & 0.68                                  \\ \hline
\multicolumn{1}{|l|}{Jirasoftware}        & JI                                   & 1.62                                   & 0.92                                 & 1.38                                  \\ \hline
\multicolumn{1}{|l|}{Mesos}               & ME                                   & 1.21                                   & 0.66                                 & 1.02                                  \\ \hline
\multicolumn{1}{|l|}{Moodle}              & MD                                   & 5.34                                   & 4.09                                 & 5.97                                  \\ \hline
\multicolumn{1}{|l|}{Mule}                & MU                                   & 2.47                                   & 1.43                                 & 2.18                                  \\ \hline
\multicolumn{1}{|l|}{Mule Studio}         & MS                                   & 3.58                                   & 2.04                                 & 3.23                                  \\ \hline
\multicolumn{1}{|l|}{Spring XD}           & XD                                   & 1.72                                   & 0.96                                 & 1.63                                  \\ \hline
\multicolumn{1}{|l|}{Talend Data Quality} & TD                                   & 2.20                                   & 1.58                                 & 2.97                                  \\ \hline
\multicolumn{1}{|l|}{Talend ESB}          & TE                                   & 0.91                                   & 0.50                                 & 0.64                                  \\ \hline
\multicolumn{1}{|l|}{Titanium}            & TI                                   & 2.04                                   & 1.36                                 & 1.97                                  \\ \hline
\multicolumn{1}{|l|}{Usergrid}            & UG                                   & 1.16                                   & 0.68                                 & 1.03                                  \\ \hline
\multicolumn{2}{|l|}{\textbf{Average}}                                                       & 2.38                                   & 1.44                                 & 2.09                                  \\ \hline
\end{tabular}
\label{tbl:RQ1}
\end{table}
\subsection{RQ2: Cross Project Within Repository  effort estimation}
The result of predicting story points for cross projects within the repository is shown in Table \ref{tbl:RQ2}. In this evaluation, we predict the data of issues for 7 software projects based on the issues provided. by another project from the same repository. For this evaluation, we outperformed GPT2SP for most of the target projects. We achieved about the same MAE as 1.5 for the prediction of the Mesos project. The best project that our tool HeteroSP outperformed GPT2SP can perform in this setting is the Usergrid project. A possible reason for the good MAE is thanks to the ratio of training and testing as 4:1, which can be significant to avoid a lack of training data. The project that caused challenges for story point prediction is the Aptana Studio project, which has the highest MAE returned by HeteroSP. We achieved consistently the MAE of prediction for the Titanium project at 3.17 and 3.18. Compared to Deep-SE, we achieved better accuracy with 3 configurations including AP-TI, ME-UG, and TI-AS.

For the training time, HeteroSP required 238 seconds to provide the training for 8 projects. Though this is a larger number compared to the running time in RQ1, it is reasonable since for RQ2 we need to run on the whole dataset of story points for several projects. HeterroSP took 0.23 seconds to get the testing result for 8 projects. 

\textbf{HeteroSP outperforms GPT2SP in predicting the SPs with lower MAE by 0.16 but got higher MAE than Deep-SE by 0.2 for cross-project within repository estimation.}

\begin{table}
\small
\centering
\caption{RQ2: Comparison of MAE Between HeteroSP and baselines for Cross Project Within Repository Estimation}
\begin{tabular}{|ll|r|r|r|}
\hline
\multicolumn{1}{|l|}{\textbf{Source}} & \textbf{Target} & \multicolumn{1}{c|}{\textbf{HeteroSP}} & \multicolumn{1}{c|}{\textbf{GPT2SP}} & \multicolumn{1}{c|}{\textbf{Deep-SE}} \\ \hline
\multicolumn{1}{|l|}{AS}              & AP              & 4.17                                   & 4.45                                 & 2.78                                  \\ \hline
\multicolumn{1}{|l|}{AS}              & TI              & 3.17                                   & 3.20                                 & 2.06                                  \\ \hline
\multicolumn{1}{|l|}{AP}              & TI              & 3.18                                   & 3.71                                 & 3.45                                  \\ \hline
\multicolumn{1}{|l|}{ME}              & UG              & 0.89                                   & 0.98                                 & 1.16                                  \\ \hline
\multicolumn{1}{|l|}{MS}              & MU              & 2.60                                   & 2.66                                 & 2.31                                  \\ \hline
\multicolumn{1}{|l|}{MU}              & MS              & 3.26                                   & 3.36                                 & 3.14                                  \\ \hline
\multicolumn{1}{|l|}{TI}              & AS              & 2.10                                   & 2.29                                 & 3.22                                  \\ \hline
\multicolumn{1}{|l|}{UG}              & ME              & 1.50                                   & 1.50                                 & 1.13                                  \\ \hline
\multicolumn{2}{|c|}{\textbf{Average}}                  & 2.61                                   & 2.77                                 & 2.41                                  \\ \hline
\end{tabular}
\label{tbl:RQ2}
\end{table}

\subsection{RQ3: Cross Project Cross Repository  effort estimation}
\begin{table*}
\small
\centering
\caption{Comparison Between HeteroSP and models from RQ3, RQ4, RQ5, RQ6 and RQ7}

\begin{tabular}{|ll|r|rr|rrr|rr|rr|r|}
\hline
\multicolumn{2}{|l|}{\textbf{Cross Proj. Cross Repo.}}  & \multicolumn{1}{l|}{}                  & \multicolumn{2}{c|}{\textbf{RQ3}}                                           & \multicolumn{3}{c|}{\textbf{RQ4}}                                                                                                                           & \multicolumn{2}{c|}{\textbf{RQ5}}                                                               & \multicolumn{2}{c|}{\textbf{RQ6}}                                           & \multicolumn{1}{c|}{\textbf{RQ7}}       \\ \hline
\multicolumn{1}{|l|}{\textbf{Source}} & \textbf{Target} & \multicolumn{1}{c|}{\textbf{HeteroSP}} & \multicolumn{1}{c|}{\textbf{GPT2SP}} & \multicolumn{1}{c|}{\textbf{Deep-SE}} & \multicolumn{1}{c|}{\textbf{AGNN}}                        & \multicolumn{1}{l|}{\textbf{GAT}}                         & \multicolumn{1}{l|}{\textbf{GCN}}    & \multicolumn{1}{c|}{\textbf{Title}}                       & \multicolumn{1}{c|}{\textbf{Desc}}   & \multicolumn{1}{c|}{\textbf{Acc.}}    & \multicolumn{1}{c|}{\textbf{MAE}}    & \multicolumn{1}{l|}{\textbf{Parse Tree}} \\ \hline
\multicolumn{1}{|l|}{\textbf{AS}}     & \textbf{MU}     & \textbf{2.60}                          & \multicolumn{1}{r|}{2.63}            & 2.70                                  & \multicolumn{1}{r|}{2.83}                                 & \multicolumn{1}{r|}{2.72}                                 & 3.03                                 & \multicolumn{1}{r|}{2.60}                                 & 2.60                                 & \multicolumn{1}{r|}{26.21\%}          & 2.60                                 & 2.61                                     \\ \hline
\multicolumn{1}{|l|}{\textbf{AS}}     & \textbf{MS}     & \textbf{3.26}                          & \multicolumn{1}{r|}{3.32}            & 4.24                                  & \multicolumn{1}{r|}{3.26}                                 & \multicolumn{1}{r|}{{\color[HTML]{FF0000} \textbf{3.25}}} & 3.26                                 & \multicolumn{1}{r|}{3.26}                                 & 3.26                                 & \multicolumn{1}{r|}{30.46\%}          & 3.26                                 & 3.26                                     \\ \hline
\multicolumn{1}{|l|}{\textbf{CV}}     & \textbf{UG}     & \textbf{0.90}                          & \multicolumn{1}{r|}{1.06}            & 1.57                                  & \multicolumn{1}{r|}{1.83}                                 & \multicolumn{1}{r|}{1.79}                                 & 1.69                                 & \multicolumn{1}{r|}{{\color[HTML]{FF0000} \textbf{0.89}}} & {\color[HTML]{FF0000} \textbf{0.89}} & \multicolumn{1}{r|}{18.05\%}          & 1.85                                 & 0.91                                     \\ \hline
\multicolumn{1}{|l|}{\textbf{MS}}     & \textbf{TI}     & \textbf{3.17}                          & \multicolumn{1}{r|}{3.25}            & 6.36                                  & \multicolumn{1}{r|}{3.17}                                 & \multicolumn{1}{r|}{3.17}                                 & 3.18                                 & \multicolumn{1}{r|}{3.17}                                 & 3.17                                 & \multicolumn{1}{r|}{31.27\%}          & 3.17                                 & 3.17                                     \\ \hline
\multicolumn{1}{|l|}{\textbf{MU}}     & \textbf{TI}     & \textbf{3.17}                          & \multicolumn{1}{r|}{3.27}            & {\color[HTML]{FF0000} \textbf{2.67}}  & \multicolumn{1}{r|}{{\color[HTML]{FF0000} \textbf{3.15}}} & \multicolumn{1}{r|}{{\color[HTML]{FF0000} \textbf{3.12}}} & 3.17                                 & \multicolumn{1}{r|}{3.17}                                 & 3.17                                 & \multicolumn{1}{r|}{31.27\%}          & 3.17                                 & 3.17                                     \\ \hline
\multicolumn{1}{|l|}{\textbf{TD}}     & \textbf{AS}     & \textbf{2.10}                          & \multicolumn{1}{r|}{2.59}            & 3.11                                  & \multicolumn{1}{r|}{2.10}                                 & \multicolumn{1}{r|}{2.40}                                 & 3.00                                 & \multicolumn{1}{r|}{2.10}                                 & 2.10                                 & \multicolumn{1}{r|}{26.17\%}          & 3.17                                 & 2.10                                     \\ \hline
\multicolumn{1}{|l|}{\textbf{TD}}     & \textbf{AP}     & \textbf{4.28}                          & \multicolumn{1}{r|}{4.96}            & 5.37                                  & \multicolumn{1}{r|}{{\color[HTML]{FF0000} \textbf{3.74}}} & \multicolumn{1}{r|}{{\color[HTML]{FF0000} \textbf{3.72}}} & {\color[HTML]{FF0000} \textbf{3.75}} & \multicolumn{1}{r|}{4.30}                                 & 4.29                                 & \multicolumn{1}{r|}{34.62\%}          & {\color[HTML]{FF0000} \textbf{3.76}} & 4.30                                     \\ \hline
\multicolumn{1}{|l|}{\textbf{TE}}     & \textbf{ME}     & \textbf{1.54}                          & \multicolumn{1}{r|}{1.72}            & 2.08                                  & \multicolumn{1}{r|}{2.09}                                 & \multicolumn{1}{r|}{2.01}                                 & {\color[HTML]{FF0000} \textbf{1.47}} & \multicolumn{1}{r|}{1.51}                                 & 1.51                                 & \multicolumn{1}{r|}{29.64\%}          & 1.50                                 & {\color[HTML]{FF0000} \textbf{1.51}}     \\ \hline
\multicolumn{2}{|c|}{\textbf{Average}}                  & \textbf{2.63}                          & \multicolumn{1}{r|}{\textbf{2.85}}   & \textbf{3.51}                         & \multicolumn{1}{r|}{\textbf{2.77}}                        & \multicolumn{1}{r|}{\textbf{2.77}}                        & \textbf{2.82}                        & \multicolumn{1}{r|}{{\color[HTML]{FF0000} \textbf{2.62}}} & {\color[HTML]{FF0000} \textbf{2.62}} & \multicolumn{1}{r|}{\textbf{28.46\%}} & \textbf{2.81}                        & \textbf{2.63}                            \\ \hline
\end{tabular}
\label{tbl:RQ3To7}
\end{table*}

For this scenario, the projects for training and projects for testing are from different repositories. The result is shown in Table \ref{tbl:RQ3To7}. Compared to Deep-SE, HeteroSP achieved a significantly better MAE of 2.63 over 3.51. We also achieved better MAE than GPT2SP in 8 target projects. With the Titanium project, in this setting, we used Mule and Mule Studio projects as the source projects for training. We got almost the same MAE with different source projects training for predicting issues of Titanium. This fact means training projects from the same repository can construct models that have the same performance for effort estimation of unseen issues. We achieved the best MAE with Clover to Usergrid prediction as 0.9. Similarly, GPT2SP achieved the best MAE in this project. While both HeteroSP and GPT2SP achieved high MAE for prediction from the Talend-Data Quality project to the Aptanta-Studio project, HeteroSP achieved better MAE with over 0.6 as the lower distance compared to GPT2SP.  

We require the running time from 17.6 seconds to 32 seconds to complete the training for each project. The total testing time to run on all projects is 0.2 seconds. 

\textbf{HeteroSP outperforms GPT2SP by 0.22 and Deep-SE by 0.88 in MAE for cross-project cross repository story point estimation.} 
\subsection{RQ4: Comparison between Heterogeneous Graph and Homogeneous Graph}
While prior work on estimating story points using Text Level Graph Neural Networks \cite{https://doi.org/10.48550/arxiv.2203.03062} confronts challenges in estimating story points, we want to test the hypothesis that newer homogeneous GNN can work with this problem. By the result shown in Table \ref{tbl:RQ3To7}, we see that. the HeteroSP outperformed three homogeneous GNN models by 0.14 to 0.19 in the average MAE of cross-project cross repository effort estimation. HeteroSP achieved significantly better MAE in the prediction of Usergrid from Clover. There are projects in that HeteroSP got higher MAE compared to other homogeneous approaches. They are Titanium and Mule Studio projects. However, overall HeteroSP achieved 2.63 of average MAE while homogeneous GNN models achieved from 2.77 to 2.82 of MAE. 

For the running time, GCN required 1.52 seconds for training and 0.001 seconds for testing, while AGNN required 4.13 seconds and 0.0026 seconds. GAT required 6.26 seconds for training and 0.003 seconds for testing. In general, they have better training time performance compared to HeteroSP, which is reasonable since they don't need to be trained on multiple types of nodes and edges.  Both GCN, AGNN, and GAT have better running time than Text-Level GNN \cite{https://doi.org/10.48550/arxiv.2203.03062}. A possible reason for the improvement of performance compared to the previous GNN approach is that in this project we design a better graph representation for the prediction. In \cite{https://doi.org/10.48550/arxiv.2203.03062}, the graph generation is provided at windows of n-grams of words as input, which caused the exponential of vocabulary for training due to the characteristics of the story point dataset.

\textbf{Heterogeneous GNN achieves better MAE than homogeneous GNN approaches but got worse time performance.}

\subsection{RQ5: Estimation of Issues lacking of title or description}
The result of prediction on issues that are incomplete is shown in Table \ref{tbl:RQ3To7}. It shows that the MAE on the configuration of predicting only title or description of software issues provides a slightly better MAE. Projects that got better MAE with lack of text are Mesos and Usergrid. In project Appcelerator, the MAE goes high with incomplete issue description but not significantly. 

\textbf{HeteroSP can solve the prediction for incomplete issues with comparable accuracy.}

\subsection{RQ6: Accuracy of Estimation's Classification}
From Table \ref{tbl:RQ3To7}, we can see that the accuracy score for software effort classification is from 26\% to over 34\% while the MAE ranges from 2.6 to 3.76. We got the average accuracy of classification as 28.446\% while the average MAE is 2.81. 

\textbf{With HeteroSP, the SP classification model achieves lower accuracy than the SP regression model.}  

\subsection{RQ7: Effects of statistical NLP parse tree to SP estimation}
From Table \ref{tbl:RQ3To7}, we observe that the average MAE if we include the parse tree of StanfordNLP in the graph is the same as the MAE when we train on our graph. The parse tree can improve the MAE for Mesos project estimation, while it had worse performance on predicting issues of the Aptana Studio project. Moreover, the time for generating the parse tree of 23313 issues in this studied dataset is over 8 hours.

\textbf{Compared to the NLP parse tree, HeteroSP achieves a comparable MAE but our algorithm requires significantly less time for graph construction.}

\section{Related Work}
\textbf{Graph Neural Networks in Software Engineering.} Machine Learning techniques on sequential textual input have been widely applied on many research problems in SE such as specification inference \cite{7966873}, type inference \cite{10.1145/3180155.3180230} and natural language to code translation \cite{10.1109/ICSE-C.2017.81,DBLP:conf/kbse/PhanS021}. Compare to these works, GNN-based models have advantages on allow graph representation of the input.  In SE, research that applies GNN models mainly focuses on the representation of the Abstract Syntax Tree (AST). Allamanis et al. \cite{DBLP:journals/corr/abs-1711-00740} optimized the Gate Graph Neural Networks to solve node-level classification problems using AST for predicting the name of the variables. Brockschmidt et al. \cite{DBLP:journals/corr/abs-1805-08490} proposed a graph extension procedure to synthesize expression from program location. Aravind et al. \cite{10.1145/3382494.3410675} propose a GNN model to detect the similarity between two input programs. Tehrani et al. \cite{https://doi.org/10.48550/arxiv.2203.00611} applied Relational Graph Convolutional Networks (RGCN) model for predicting  optimum NUMA/prefetcher configurations of  C/C++ programs. In future, applying techniques on optimization of GNN models \cite{yu2021auto,yu2021gnnrl} can help increasing the performance and accuracy in solving SE problems.

\textbf{Heterogeneous Graph Transformer Models.} The idea of the HGT model is proposed by the work of Hu et al. \cite{DBLP:journals/corr/abs-2003-01332}. Their HGT model was originally built to support web-scale graph datasets. Fey et al. \cite{fey2019fast} implement a library that included multiple GNN models including the HGT transformer developed by Hu et al. \cite{DBLP:journals/corr/abs-2003-01332}. These works formulate the research problem as node-level classification. In SE, Wang et al. \cite{DBLP:journals/corr/abs-2012-04188} design a model for method name generation using heterogeneous graph. They adjust the default HGT model for classification to an encoder-decoder layer for text summarization. 

\textbf{Other Approaches on SPE.} In this paper, we already mentioned Porru's approach \cite{10.1145/2972958.2972959}, Deep-SE \cite{DBLP:journals/corr/ChoetkiertikulD16} and GPT2SP \cite{9732669} are three well-known works for story point estimation. There are some other works related to this area that have appeared recently. Phan et al. \cite{https://doi.org/10.48550/arxiv.2203.03062} studied problems of the GNN model for this SE problem. Morais et al. \cite{de2021deep} apply multiple deep learning algorithms on the Deep-SE dataset. They also do the preprocessing on input text but achieved worse accuracy compared to Deep-SE \cite{DBLP:journals/corr/ChoetkiertikulD16} due to the ineffective of their deep learning models. To the best of our knowledge, HeteroSP the first work that fully implements and optimizes GNN models with competitive accuracy to achieve this task.

\textbf{A recent study on Deep-SE and future directions for SPE.} In June 2022 (after the time our paper was submitted to ESEM 2022), Tawosi et al. \cite{DBLP:journals/corr/abs-2201-05401} conducted a replication study on Deep-SE and compared Deep-SE over the traditional text regression approach using Term Frequency Inverse Document Frequency (TFIDF) for input feature extraction, called TFIDF-SE. They had some findings and conclusions. First, the replicated implementation of Deep-SE provided by Tawosi et al received lower accuracy than the results reported in the Deep-SE paper \cite{DBLP:journals/corr/ChoetkiertikulD16} in both three configurations (RQs 1-3). They concluded that possible reasons for these discrepancies were the incorrectness of MAE computation in the original work of Deep-SE proposed by Choetkiertikul et al. \cite{DBLP:journals/corr/ChoetkiertikulD16} along with the private ways that Deep-SE used for random guessing and splitting the train/valid/test data. Second, Tawosi et al. showed that the re-implemented version of Deep-SE didn't statistically outperform the classical TFIDF-SE approach in most cases. This fact suggests future works on SP estimation should compare their works carefully with traditional approaches and use other metrics which can be more meaningful than MAE. Third, they provide some new directions for researching in this area. Similar to our RQ6, they suggest solving the problem by multi-label classification models can be a more suitable direction compared to regression techniques. Moreover, they conducted a new dataset of software issues story points \cite{DBLP:journals/corr/abs-2202-00979}, which contains over 39000 entities and can be a good substitution for the Deep-SE dataset in future works.

\section{Threats To Validity}
There are several threats to validity that we want to discuss. First, the dataset of story points proposed by Deep-SE \cite{DBLP:journals/corr/ChoetkiertikulD16} might not be representative. We mitigate this threat by a comprehensive literature review and see that most current-trend tools for story point estimation used this dataset for evaluation. Second, there can be a risk that the task of estimating story points is a trivial task that does not require models for automatic inference. GPT2SP \cite{9732669} conducted a study on a group of software developers and the study shows that story point estimation is a challenging task even with senior developers. Third, there can be an alternative graph generation tool from natural language description for RQ7 such as NLTK Bllip parser \cite{urlPOSParseTree} and POSIT \cite{10.1145/3377811.3380440}. We already checked these tools. NLTK Bllip parser hasn't been updated for more than 7 years and POSIT didn't support tree and dependency edges generation. We conclude that StanfordCoreNLP \cite{manning2014stanford} is the best tool for our experiment. 
\section{Conclusion}
In this work, we propose HeteroSP, a framework for predicting story points from the software issues dataset proposed by Choetkiertikul et al \cite{DBLP:journals/corr/ChoetkiertikulD16}. To overcome the problems of large size vocabulary and the nature of issues' descriptions which are mixed of text and code, we provide an approach to representing issues' descriptions by a heterogeneous graph. By the evaluation, we show that HeteroSP performs best with cross-project cross repository estimation over GPT2SP and Deep-SE. Moreover, our approach requires less than 600 seconds for three steps: initial node embedding, graph generation, and model construction, which is much more efficient than deep learning-based approaches like Deep-SE \cite{DBLP:journals/corr/ChoetkiertikulD16}. In future research, we will work on improving the performance of our model by other strategies of graph representation and text normalization. Another direction for this research problem is to integrate other information along with textual information on software issues, such as the source code of software repositories. 
\section*{Acknowledments}
We acknowledge the authors of GPT2SP and Deep-SE for providing replication packages. We thank Instructor Mouly Kumar from the Department of Electrical and Computer Engineering and Dr. Theresa Windus from the Department of Chemistry of Iowa State University (ISU) for hiring and providing funding support for the first author. We thank Ms. Nicole Lewis and the Department of Computer Science of ISU for providing a working environment for this project. We are grateful to Dr. Frank Proschan, an alumnus of the University of Texas at Austin, Dr. Susan Bayly from the University of Cambridge, and some other friends of the first author for carefully proofreading and revising the writing of this paper. 
\clearpage
\bibliographystyle{ACM-Reference-Format}
\bibliography{references}

\end{document}